\documentclass[aps,prl,twocolumn,superscriptaddress,showpacs,amsmath,amssymb]{revtex4}
\usepackage{graphicx}
\usepackage{epstopdf}
\usepackage{longtable}
\usepackage{hyperref}
\usepackage{amssymb}

\usepackage{bm}

\begin{document}

\title{Exotic Magnetism and Spin-Orbit Assisted Mott Insulating State in a $3d-5d$ Double Perovskite}

\author{A. S. Cavichini}
\affiliation{Universidade Federal do Esp\'irito Santo, Vit\'oria, Esp\'irito Santo 29075-910, Brazil}

\author{M. T. Orlando}
\affiliation{Universidade Federal do Esp\'irito Santo, Vit\'oria, Esp\'irito Santo 29075-910, Brazil}

\author{J. B. Depianti}
\affiliation{Universidade Federal do Esp\'irito Santo, Vit\'oria, Esp\'irito Santo 29075-910, Brazil}

\author{J. L. Passamai Jr.}
\affiliation{Universidade Federal do Esp\'irito Santo, Vit\'oria, Esp\'irito Santo 29075-910, Brazil}

\author{F. Damay}
\affiliation{Laboratoire L\'eon Brillouin, CEA, Centre National de la Recherche Scientifique, CE-Saclay, 91191 Gif-sur-Yvette, France}

\author {F. Porcher}
\affiliation{Laboratoire L\'eon Brillouin, CEA, Centre National de la Recherche Scientifique, CE-Saclay, 91191 Gif-sur-Yvette, France}

\author{E. Granado}
\affiliation{``Gleb Wataghin'' Institute of Physics, University of Campinas - UNICAMP, Campinas, S\~ao Paulo 13083-859, Brazil}

\begin{abstract}

The magnetic structure of Ca$_2$MnReO$_6$ double perovskite is investigated by neutron powder diffraction and bulk magnetization, showing dominant non-collinear Mn magnetic moments [$4.35(7)$ $\mu_B$] that are orthogonally aligned with the small Re moments [$0.22(4)$ $\mu_B$]. {\it Ab}-initio electronic structure calculations show that the strong spin-orbit coupling for Re $5d$ electrons combined with a relatively modest on-site Coulomb repulsion ($U_{eff}^{Re} \gtrsim 0.6$ eV) is sufficient to render this material insulating. This is a rare example of spin-orbit assisted Mott insulator outside the realm of iridates, with remarkable magnetic properties.

\end{abstract}

\maketitle

The identification of a spin-orbit entangled Mott insulating state in Sr$_2$IrO$_4$ triggered a quest for novel quantum phases in $5d$-based materials \cite{Kim,Kim2,Pesin,Caorev,Jackeli}. In such systems with strong spin-orbit coupling (SOC), a Heisenberg Hamiltonian is not sufficient to describe the magnetic ground states and excitations, therefore alternative treatments may be necessary to describe the effect of anisotropic exchange interactions \cite{Jackeli}. Also, the electronic correlations in $5d$ systems are relatively weak, and the atomic moments are normally small compared to the $3d$ counterparts. Alternating $3d$ and $5d$ ions in an ordered double perovskite structure offers a possible pathway to investigate ground states arising from the combination of strong SOC in $5d$ ions and strong electronic correlation in $3d$ ions. An example is Ca$_2$MnReO$_6$ (CMRO), which was previously described as a ferromagnetic insulator with a small saturation magnetization of 0.9 $\mu_B/$f.u. and coercive field of 4 T \cite{Kato}, also showing Mn valence close to +2 \cite{Correa,Depianti}. Here, an unusual magnetic structure is demonstrated, with dominating Mn moments forming a largely canted sublattice, while weak Re moments are orthogonally aligned with Mn moments. Electronic structure calculations reveal that a combination of Re SOC and electronic correlations renders this material insulating, revealing a determinant role of the Re moments to the overall electronic and magnetic behavior of this material. 

The polycrystalline CMRO sample was synthesized by solid state reaction \cite{Depianti}. dc-magnetization and magnetic susceptibility measurements were performed in a commercial platform. Cold neutron powder diffraction (c-NPD) measurements were taken at the $G$4-1 instrument of Laboratorie L\'eon Brillouin (LLB), using a highly oriented pyrolytic graphite (HOPG) monochromator with vertical focusing and $\lambda = 2.43$ \AA, a two-axis diffractometer and a BF$_3$ multicell detector with $80^{\circ}$ aperture and $0.1^{\circ}$ resolution. High resolution thermal neutron powder diffraction (t-NPD) measurements were also taken, and details are given in the Supplemental Material (SM) \cite{supplemental}. The sample was kept sealed under He atmosphere in a vanadium can and mounted into the cold finger of a LHe cryostat. The powder diffraction data were independently analysed using the Fullprof \cite{Fullprof} and GSAS \cite{GSAS} suites, with similar results. The refined degree of Mn/Re antisite disorder is 0.6(7) \%. The Re$^{6+}$ magnetic form factor was taken from Ref. \cite{Popov}. {\it Ab}-initio electronic structure calculations were carried out with relativistic Density-Functional Theory using the QUANTUM ESPRESSO package \cite{QE} under the Generalized Gradient Approximation, using the Perdew, Burke, and Enzerhof exchange-correlation potential \cite{Perdew} and the Projector Augmented Wave method \cite{PAW,pseudopotentials}. The energy cuttoffs for the wavefunctions and charge density were 57 and 702 Ry, respectively. The atomic-projected magnetizations were obtained by a self-consistent field (scf) calculation using a $6 \times 6 \times 4$ Monkhorst-Pack grid of $k$-points. The density of states was obtained from a subsequent non-scf calculation with a denser $12 \times 12 \times 8$ $k$-point grid, using 0.03 eV Gaussian broadening. The experimental crystal structure was employed as input of our calculations.

\begin{figure}
\includegraphics[width=0.4\textwidth]{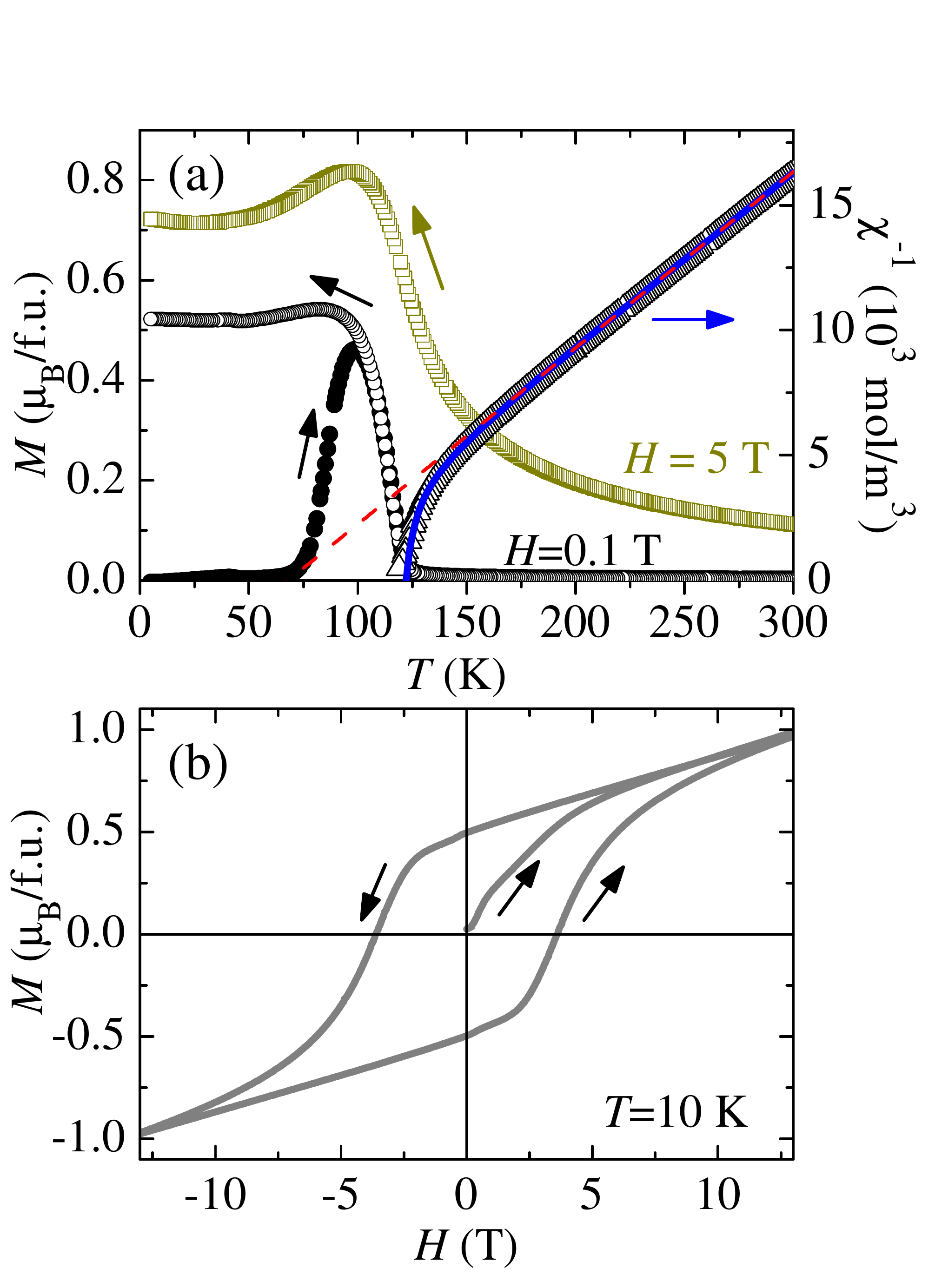}
\caption{\label{Mag} (color online) (a) $T$-dependence of the dc-magnetization of Ca$_2$MnReO$_6$ under warming at $H=0.1$ T after zero-field cooling (ZFC, filled circles), field cooling (FC) at $H=0.1$ T (empty circles) and FC at $H=5$ T (empty squares). The inverse magnetic susceptibility obtained from FC magnetization data at $H=0.1$ T is shown as empty triangles, together with fits to the Curie-Weiss law (dashed line) and the two-sublattice paramagnetic model (solid line, see text).}
\end{figure}

Figure \ref{Mag}(a) shows $M(T)$ curves of CMRO for $H= 0.1$ and 5 T. The zero-field cooled (ZFC) magnetization with $H=0.1$ T is negligible up to 70 K, shows a peak at $T_{max}=100$ K and a paramagnetic transition at $T_c = 121$ K. The $H= 0.1$ T field-cooling (FC) curve follows the ZFC one above $T_{max}$, retaining a significant magnetization ($\sim 0.5$ $\mu_B$/f.u.) below $T_{max}$. The $H=5$ T FC magnetization also shows a transition at $T_c$ and a peak at $T_{max}$, reaching $M = 0.7$ $\mu_B$ at the base $T$. The inverse magnetic susceptibility $\chi^{-1}$ taken with $H=0.1$ T, also shown in Fig. \ref{Mag}(a), follows a Curie-Weiss law in the high-$T$ limit, $\chi^{-1}= (T-\theta_{CW})/C$ with $\theta_{CW}=70$ K and $C=1.42 \times 10^{-2}$ Km$^{3}$/mol, with relevant deviations from this law below $\sim 150$ K [see Fig. \ref{Mag}(a)]. The positive value of $\theta_{CW}$ indicates mostly ferromagnetic exchange interactions between the dominant magnetic moments. Alternatively, a fit to a two-sublattice paramagnetic model, $\chi^{-1}= T/C_m + 1/\chi_0 -b/(T-\theta)$ (Ref. \cite{Cullity}), yields excellent agreement with the data above $T_c$, with fitting parameters $C_m = 1.45\times10^{-2}$ K m$^{3}$/mol , $\chi_0 = 2.36\times10^{-4}$ m$^{3}$/mol, $b = 1.84\times10{^4}$ K.mol/m$^3$ and $\theta = 118$ K.  The $M \times H$ loop taken at $T=10$ K after ZFC is given in  Fig. \ref{Mag}(b), showing a squared curve with large coercive field $H_{c}=4$ T, characteristic of significant magnetocrystalline coupling, and remnant magnetization $M_{rem} = 0.5$ $\mu_B$/f.u., in line with Ref. \cite{Kato}.

\begin{figure}
\includegraphics[width=0.4\textwidth]{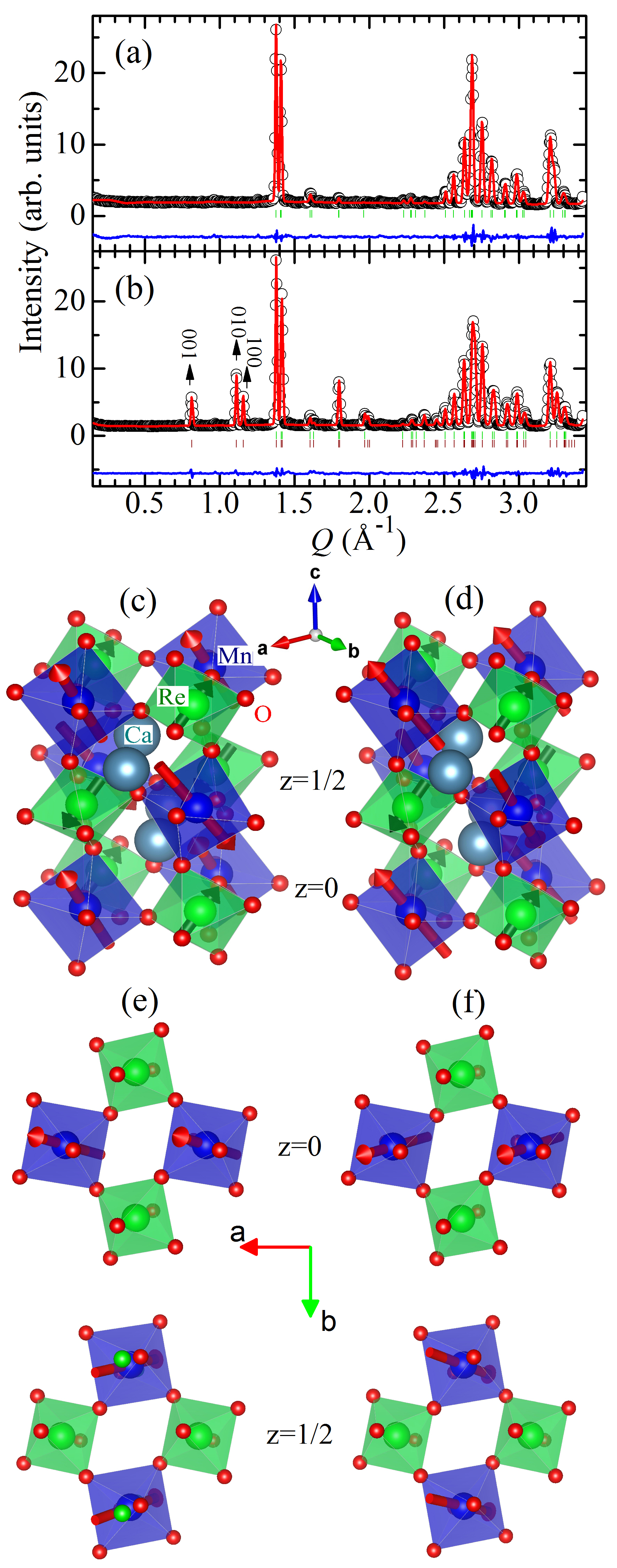}
\caption{\label{profiles} (color online) Observed cold neutron powder diffraction profile of CMRO at $T=300$ K (a) and $T=2$ K (b) (empty circles). The calculated profiles for a monoclinic double perovskite structure and non-collinear Mn and Re magnetic moments at 2 K (see text) are shown as solid lines. The difference curves (dark solid line) are displayed at the bottom of each figure. The reflection positions are indicated by short vertical lines for both the nuclear (a and b) and magnetic (b, lower vertical lines) structures. (c,d) Representations of the two possible magnetic structures at $T=2$ K (arrows), yielding identical magnetic Bragg intensities. The magnitude of the weak Re moment is exaggerated for clarity. (e,f) {\it ab}-plane projections of the representations shown in (c,d). Re moments are not displayed here. Note that in the structure of (c) and (e) the Mn moments follow the tilting pattern of the MnO$_6$ octahedra, contrary to the structure shown in (d) and (f).}
\end{figure}

\begin{figure}
\includegraphics[width=0.4\textwidth]{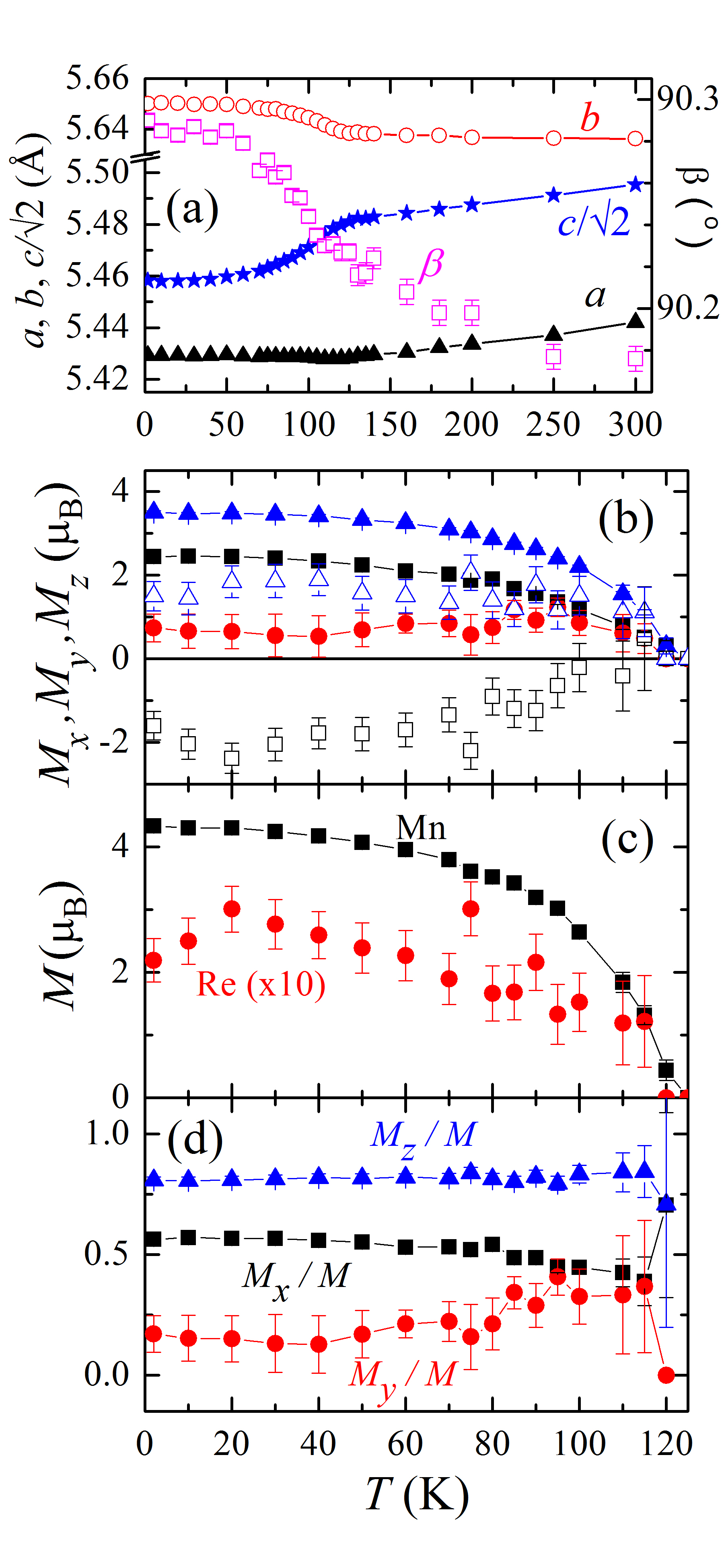}
\caption{\label{Tdep} (color online) $T$-dependence of (a) lattice parameters, (b) $M_x$(Mn) (filled squares), $M_y$(Mn) (filled circles), $M_z$(Mn) (filled triangles), $M_x$(Re) $\times 10$ (empty squares), and $M_z$(Re) $\times 10$ (open triangles), (c) total Mn and Re moments, and (d) direction cosines $M_i/M$ of the Mn moments.}
\end{figure}

Figure \ref{profiles}(a) shows the c-NPD profile of CMRO at $T=300$ K. t-NPD data are shown in the SM \cite{supplemental}. The intensities could be well modeled by a monoclinic double perovskite structure with space group $P2_1/n$, as indicated by the solid lines of Fig. \ref{profiles}(a). Refined structural parameters at selected $T$, using t-NPD data, are also given in the SM \cite{supplemental}. The $T$-dependence of the refined lattice parameters is shown in Fig. \ref{Tdep}(a). An anomalous contraction of $c$ and expansion of $b$ are observed on cooling below $T_c$.

Figure \ref{profiles}(b) shows the c-NPD profile of CMRO at $T=2$ K. Additional intensities due to magnetic ordering are observed at integer $(hkl)$ positions, showing that the magnetic unit cell coincides with the chemical one. In principle, the magnetic (Shubnikov) groups $P2_1/n$ and $P2_1^{'}/n'$ are allowed with non-zero magnetic moments on Mn or Re sites. The former corresponds to an $A$-type antiferromagnetic ($A$-AFM) structure with the magnetic moments in the {\it ac} plane with a possible ferromagnetic (FM) component for the moments along the monoclinic principal axis {\it b}, while for the primed group the FM component is within the {\it ac} plane and the $A$-AFM moments are oriented along {\it b}. The significant monoclinic distortion of CMRO allows for an unambiguous determination of the magnetic group, in contrast to the previously studied case of Sr$_2$MnReO$_6$ with a quasi-cubic cell \cite{Popov,Popov2}. In fact, the presence of a strong (010) magnetic reflection, clearly separated from (100) [see Fig. \ref{profiles}(b)], shows that the $A$-AFM moments are not oriented along $b$, and therefore the $P2_1/n$ magnetic group is the correct one for CMRO. Still, two possible magnetic structures provide exactly the same magnetic Bragg intensities. Formally, they differ by reversing the sign of the small FM component $M_y$ (Mn). In the magnetic structure of Figs. \ref{profiles}(c,e), the canted Mn moments follow the tilts of the MnO$_6$ octahedra around the $c$-axis, always pointing approximately towards an octahedral face, while in Figs. \ref{profiles}(d,f) the canting of the Mn moments is off-phase with respect to the MnO$_6$ tilting pattern. At $T=2$ K, the Mn/Re moment vectors are $[2.41(3),\pm 0.8(3),3.53(3)]$ / $[-0.16(4),0,0.15(4)]$ $\mu_B$ at $z=0$ in Figs. \ref{profiles}(c) and \ref{profiles}(d), yielding total Mn/Re atomic moments of $4.35(7)/0.22(4)$ $\mu_B$. The refined $M_y$ (Mn) is consistent with the FM moment found in bulk magnetization measurements (see Fig. \ref{Mag}). Due to the weakness of the Re moment, its component along {\it b}, i.e., its minor contribution to the already small overall FM signal, could not be obtained, being fixed at zero. Figure \ref{profiles}(b) shows the fitting of c-NPD data to the model described above.

Figure \ref{Tdep}(b) shows the $T$-dependence of the Mn and Re moment components, while the total moments are given in Fig. \ref{Tdep}(c). Both Mn and Re total moments decrease on warming, critically approaching zero at $T_c=121$ K, consistent with bulk magnetization data. Notice that $M_y$(Mn) increases slightly between $\sim 80$ and 100 K, reaching $M_y$(Mn)$=1.2(2)$ $\mu_B$ at 95 K, consistently with the magnetization maximum found in this $T$ range [see Fig. \ref{Mag}(a)]. This effect is analyzed in term of direction cosines [see Fig. \ref{Tdep}(d)], where the relative weight of the FM component $M_y/M$(Mn) increases above 80 K at expense of $M_x/M$, indicating a moment rotation around the {\it c} axis as $T \rightarrow T_c$. 

\begin{figure}
\includegraphics[width=0.45\textwidth]{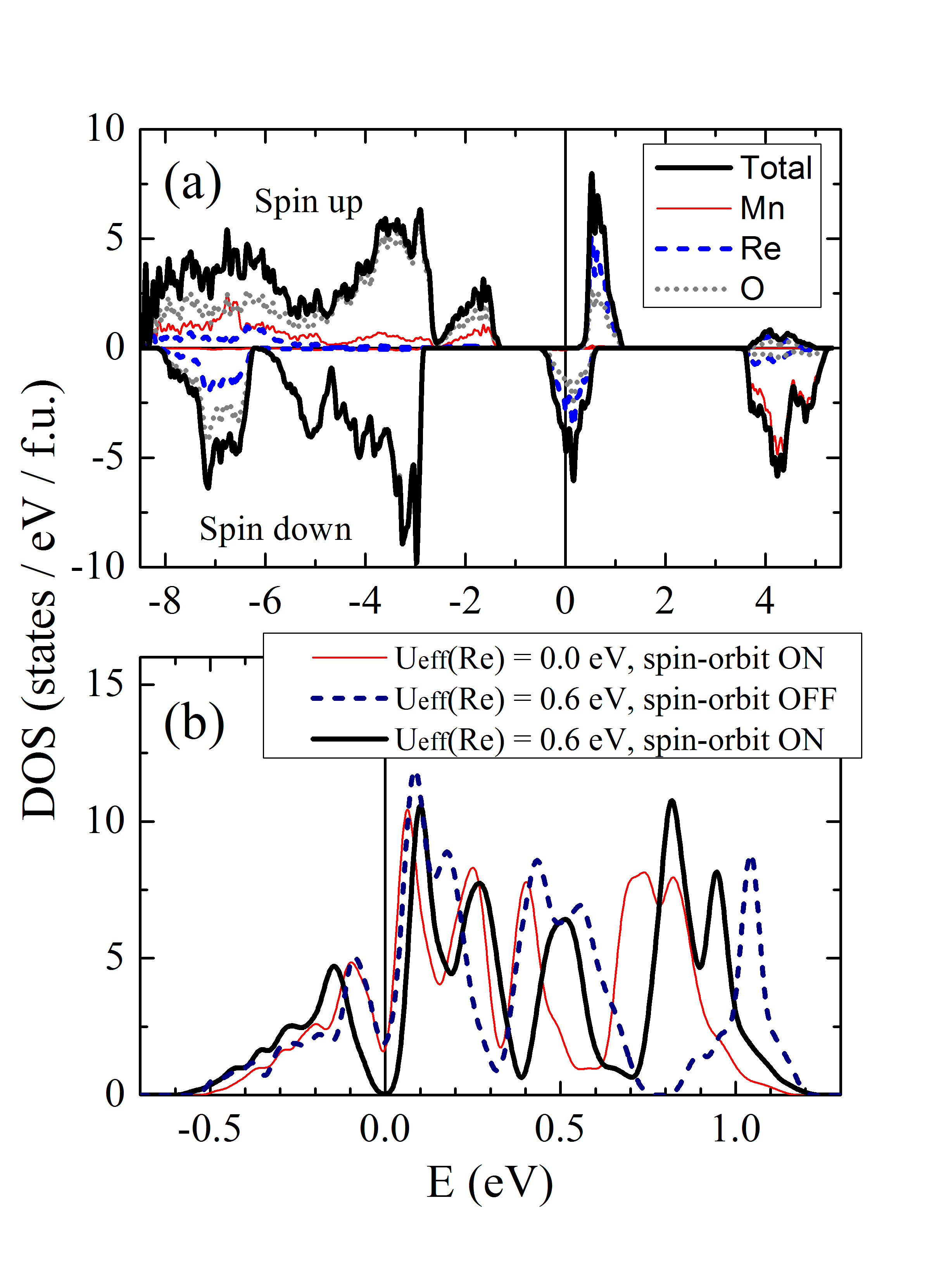}
\caption{\label{abinitio} (color online) (a) Mn $3d$, Re $5d$, O $2p$ and total density of states (DOS), taken for a collinear calculation without Re spin-orbit coupling and using $U$(Mn) = 6.9 eV, $J$(Mn) = 0.86 eV, $U$(Re)$=J$(Re)$=0$, yielding a ferrimagnetic ground state. Positive and negative values refer to spin up and spin down DOS, respectively. (b) Total DOS nearby the Fermi level for calculations with and without Re spin-orbit coupling, and with and without an effective on-site Coulomb repulsion $U_{eff}$(Re) = 0.6 eV ($U$(Re) = 1.3 eV, $J$(Re) = 0.7 eV). The DOS for the collinear calculation (spin-orbit OFF) in (b) correspond to an $A$-AFM metastable state for the Mn moments.}
\end{figure}

The magnetic structures displayed in Figs. \ref{profiles}(c) and \ref{profiles}(d) are unusual. Firstly, they show a canting of 21$^{\circ}$ of the Mn sublattice with respect to a collinear $A$-AFM structure, while the Re moments make an angle of 100(7)$^{\circ}$ with respect to the Mn moments at the same $z$. An attempt to force a collinear magnetic alignment for Mn and Re atoms led to substantially worse fitting quality, and the Re moment converged to zero within statistical error. This analysis reinforced the conclusion that a nearly orthogonal alignment of Mn and Re ions takes place, which is suggestive of dominant antisymmetric (Dzyaloshinskii-Moriya) exchange and/or single ion anisotropy energy contributions in comparison to the symmetric (Heisenberg) exchange that is normally dominant in $3d$ transition metal oxides. This is surprising considering the relatively large value of the magnetic ordering temperature of this material ($T_c=121$ K), while antisymmetric exchange coupling is normally in the range of 1 K in $3d$ transition-metal compounds. Comparable results with orthogonal Mn and Re moment alignment were reported for the related compound Sr$_2$MnReO$_6$ \cite{Popov2,Popov}. However, ferrimagnetic (FiM) contributions to the Bragg peaks are dominant over the $A$-AFM ones for Sr$_2$MnReO$_6$ \cite{Popov}, in opposition to our results for CMRO. Another analog material is the pseudo-cubic Ba$_2$MnReO$_6$, in which Mn moments show a non-collinear structure with a dominant FM component and a canting of 25$^{\circ}$ between successive (001) planes \cite{Khattak}. This indicates that compounds of this family consistently show largely non-collinear magnetic structures.

To gain insight into the origin of the exotic magnetic structure of CMRO, {\it ab}-initio electronic structure calculations were performed. Figure \ref{abinitio}(a) shows the total and atom-projected density of states near the Fermi level ($E=0$) given by a preliminary calculation with collinear magnetic moments using Coulomb ($U$) and exchange ($J$) parameters $U$(Mn) = 6.9 eV, $J$(Mn) = 0.86 eV (Ref. \cite{Anisimov}), $U$(Re)$=J$(Re)$=0$, and without considering SOC. With such parameters, the ground state is found to be half-metallic with FiM ordering of Mn and Re moments. The energy levels at the vicinity of $E=0$ show mostly Re $5d$ character, with significant mixing with O $2p$ levels but negligible participation of Mn $3d$ states. An alternative calculation performed with initial atomic magnetization defining an $A$-AFM structure converged into a metastable $A$-AFM state, with higher energy with respect to the FiM ground state ($E_{AFM}-E_{FiM}=51$ meV/f.u.). The energy difference between the $A$-AFM and FiM states decreases by increasing the Re on-site Coulomb parameters, with a crossover to an $A$-AFM ground state for $U_{eff}$(Re)$> 2.2$ eV. Inclusion of Re SOC has a large impact in the electronic structure near $E=0$. Figure \ref{abinitio}(b) shows the total DOS at the vicinity of $E=0$, with and without Re SOC and $5d$ on-site correlation. Remarkably, a relatively weak Re correlation [$U_{eff}$(Re)=0.6 eV] in the presence of Re SOC is sufficient to open a gap in the DOS at $E=0$, rendering the material insulating. Instead, if SOC is not considered, the energy gap is not formed for realistic values of $U_{eff}$(Re) $< 1.5$ eV. Therefore, this material can be classified as a spin-orbit-assisted Mott insulator, in similarity to Sr$_2$IrO$_4$ and related Ir$^{4+}$ oxides \cite{Kim,Kim2,Pesin,Caorev,Jackeli}. The atomic Mn/Re moments for the calculation including SOC and Re correlation are $4.17/0.23$ $\mu_B$, in good agreement with the experimental values. Concerning the moment directions, the calculated Re moments converged to an anti-parallel alignment with respect to the Mn moments for atoms in the same (001) plane, in contrast to the experimentally observed magnetic structure.

Both the experimental data and {\it ab}-initio calculations indicate a dominating weight of the Mn moments in CMRO. On the other hand, in an ionic picture the Mn$^{2+}$ ions with half-filled $3d$ band are isotropic with zero orbital magnetic moments, making the observed canted magnetic structure and magnetic hardness of this material difficult to interpret based upon the Mn moments alone. Thus, the Re $5d$ moments, albeit very weak, must have an important role in the magnetic properties of this material. In fact, the large Re-O hybridization may produce significant orbital polarization also in the oxygen ions that mediate the Mn-Re and Mn-Mn exchange interactions. These ingredients lead to an unusual magnetic structure for this material that cannot be trivially rationalized in terms of paradigms built upon traditional $3d$ transition-metal magnetism with much weaker effects of relativistic SOC. Further theoretical and experimental investigations are necessary for a deeper understanding of this remarkable behavior.

In summary, neutron diffraction on Ca$_2$MnReO$_6$ reveals a non-collinear magnetic structure where the Mn moments are largely canted (21$^{\circ}$) and the small Re moments are almost orthogonally aligned with the Mn ones. The large SOC of Re $5d$ electrons is identified as a decisive ingredient leading to the insulating state and the exotic magnetic structure of this material.

LLB is acknowledged for concession of beamtime. This work was supported by Fapesp, CAPES and CNPq, Brazil.


\begin{appendix}

\section{Supplemental material: Analysis of thermal neutron powder diffraction data}

\begin{figure*}
\includegraphics[width=0.9\textwidth]{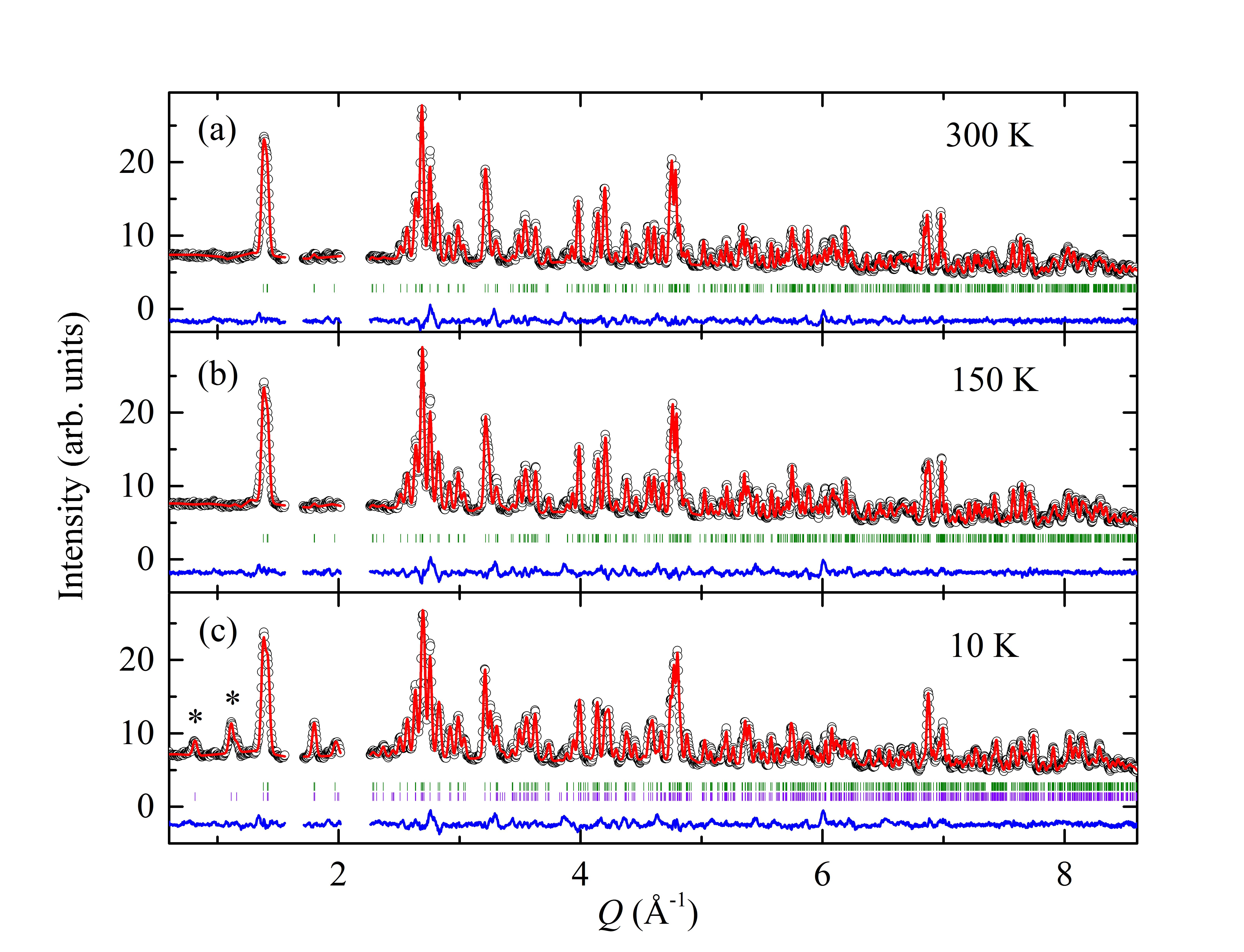}
\caption{\label{pattern} Experimental (symbols) and calculated (red lines) thermal neutron powder diffraction profile of Ca$_2$MnReO$_6$ at 300 K (a), 150K (b) and 10 K (c). The difference curves (blue solid line) are displayed at the bottom of each figure. The reflection positions are indicated in short vertical lines for both the nuclear (a, b and c) and magnetic (c, lower vertical lines) structures. Stars represent purely magnetic peaks.}
\end{figure*}

High resolution thermal neutron powder diffraction (t-NPD) measurements were taken at selected temperatures at the $3T2$ instrument of LLB, adequate for crystal structure investigations, using a vertically focusing Ge(335) monochromator with $\lambda = 1.2292$ \AA\ and a bank of 50 $^{3}$He detectors (see Fig. \ref{pattern}). We employed $10 ^{\prime}$ in-pile collimation upstream the monochromator and $10 ^{\prime}$ Soller collimators in front of the 50 counters. Contrary to the c-NPD data displayed in the main text, which were collected $\sim 6$ months after the synthesis with the sample being kept in an inert He atmosphere during this time interval, the t-NPD data were only taken after two years of sample exposure to ambient atmosphere. As an unwanted consequence of this time lapse between the experiments, unidentified impurity peaks were clearly observed at 1.62, 2.07 and 2.21 \AA$^{-1}$ in the t-NPD data, which likely arise from grain boundary carbonation. Still, satisfactory fitting quality was achieved when these impurities peaks were excluded from the refinement. The refined lattice parameters of the main phase are consistent with those obtained from the c-NPD data and the refined Mn/Re intersite disorder for the t-NPD data is $2.9(5)$ \%, slightly larger than $0.6(7)$ \% obtained from the c-NPD data, indicating only minor deterioration of the ordered double perovskite main phase in the time lapse between these experiments. Table \ref{struct} shows the refined atomic parameters at the investigated temperatures. The extracted (Mn,Re)-O bond lengths and Mn-O-Re bond angles are shown in Table \ref{bonds}.

\begingroup
\begin{table*}
\caption{\label{struct} Lattice parameters, atomic coordinates and thermal factors of Ca$_2$MnReO$_6$ in three distinct temperatures using thermal neutron powder diffraction data (space group $P2_1/n$). The $x$, $y$, and $z$ coordinates for Mn and Re are 1/2, 0, 1/2 and 1/2, 0, 0, respectively. The fitting $R$-factors are also given.}
\begin{ruledtabular}
\begin{tabular}{c c c c c} 
& 10 K & 150 K & 300 K \\
\hline
$a$ ({\AA}) & 5.4250(10) & 5.4238(9) & 5.4336(10) \\
$b$ ({\AA}) &5.6448(10) &5.6308(10) & 5.6285(10) \\
$c$ ({\AA}) & 7.7165(14) & 7.7473(14) & 7.7610(14) \\
$\beta$ ({\AA}) & 90.2816(15) & 90.2176(12) & 90.1744(12) \\
Cell Vol. {(\AA$^3$}) & 236.30(7)  & 236.60(7) & 237.35(7) \\
\hline
Ca  \\
x & 0.4873(3) & 0.4879(8) & 0.4887(8) \\
y & 0.5544(6) & 0.5541(6) & 0.5534(7) \\
z & 0.2531(5) &0.2542(5) & 0.2531(6) \\
B ({\AA$^2$}) & 0.29(1) & 0.46(1) & 0.8(1) \\
Mn/Re  \\
B ({\AA$^2$}) &0.06(5) & 0.01(4) & 0.19(4) \\
O1  \\
x & 0.3182(6) & 0.3193(6) &0.3190(6) \\
y & 0.2811(6) & 0.2819(6) & 0.2817(6) \\
z & 0.0563(4) & 0.0549(4) &0.0532(5) \\
B ({\AA$^2$}) & 0.82(4) & 0.93(4) & 1.07(4) \\
O2 \\
x & 0.2112(6) &0.2114(6) & 0.2130(6) \\
y & 0.8119(6) & 0.8135(5) & 0.8129(6) \\
z & 0.0484(4) & 0.0480(4)) &0.0474(5) \\
B ({\AA$^2$}) & 0.82(4) & 0.93(4) & 1.07(4) \\
O3  \\
x & 0.5985(6) &0.5973(6) & 0.5974(6) \\
y & -0.0312(6) & -0.0321(6) & -0.0326(6) \\
z & 0.2322(4) & 0.2348(4) & 0.2350(4) \\
B ({\AA$^2$}) & 0.82(4) & 0.93(4) & 1.07(4) \\
\hline
$R_p$ (\%) & 4.17 & 3.83 & 3.60 \\
$R_{wp}$ (\%) & 3.18 & 2.88 & 2.78 \\
$\chi^2$ & 6.01 & 5.07 & 4.31 \\

\end{tabular}
\end{ruledtabular}
\end{table*}
\endgroup

\begin{table*}
\caption{\label{bonds} (Mn,Re)-O bond distances and Mn-O-Re bond angles extracted from the data of Table \ref{struct}.}
\begin{ruledtabular}
\begin{tabular}{c c c c c} 
& 10 K & 150 K & 300 K \\
\hline
Mn-O$_1$ ({\AA}) & 2.165(4) & 2.164(4) & 2.163(4) \\
Mn-O$_2$ ({\AA}) & 2.133(3) & 2.137(3) & 2.131(4) \\
Mn-O$_3$ ({\AA}) &2.144(4) &2.131(4) & 2.133(4) \\
Re-O$_1$ ({\AA}) & 1.919(4) & 1.914(4) & 1.912(4) \\
Re-O$_2$ ({\AA}) & 1.931(4) & 1.922(4) & 1.918(4) \\
Re-O$_3$ ({\AA}) & 1.876(4) & 1.901(4) & 1.906(4) \\
Mn-O$_1$-Re ($^{\circ}$) & 146.77(18) & 146.82(18) & 147.4(2) \\
Mn-O$_2$-Re ($^{\circ}$) & 148.87(17) & 148.69(17) & 149.20(18) \\
Mn-O$_3$-Re ($^{\circ}$) & 147.33(18) & 147.76(18) & 147.69(18) \\

\end{tabular}
\end{ruledtabular}
\end{table*}

\end{appendix}

\end{document}